\documentclass[pra,showpacs,showkeys,preprint,aps]{revtex4}
\usepackage{graphicx}

\begin{document}
                              
\title{STATIC AND DYNAMIC REGIMES OF ARBITRARY GAIN COMPENSATION SINGLE-MODE LASER DIODES}

\author{C. Tannous}
\affiliation{Laboratoire de Magn\'{e}tisme de Bretagne, UPRES A CNRS 6135,
Universit\'{e} de Bretagne Occidentale, BP: 809 Brest CEDEX, 29285 FRANCE}

\date{April 14, 2001}


\begin{abstract}
We report on a methodology for the evaluation of
the DC characteristics, small-signal frequency response and
large-signal dynamic response of carrier and photon density
responses in semiconductor laser diodes. A single mode
laser is considered and described with a pair of rate equations containing
a novel non-linear gain compensation term depending on a
single parameter that can be chosen arbitrarily. This approach
can be applied to any type of solid-state laser as long as it is
described by a set of rate equations.\\


\end{abstract}

\maketitle

\section{Introduction}
Generally, lasers are described with a set of rate equations describing
the generation-recombination of carriers or emission-absorption
of photons. The equations are typically systems of first-order differential 
equations belonging to the population evolution type.\\

In this report we specialize to the simple case of single-mode laser diode where
the system of equations reduces to two: one for the carrier population (density)
and another for the photon population  (density). We concentrate on the case 
the equations embody a novel non-linear gain compensation term that depends  on a
single parameter that can be chosen arbitrarily. The effect of this parameter is studied
on the  DC characteristics, small-signal frequency response and
large-signal dynamic response of the laser. \\

The Single Mode (SM) laser diode rate equations are written as:
\begin{eqnarray}
\frac{dN}{dt} & = & \frac{I}{qV}-\frac{N}{\tau_{n}}-g(N-N_{t}) \frac{S}{{(1+\epsilon S)}^b} \\
\frac{dS}{dt} & = & \frac{\Gamma \beta N}{\tau_{n}}-\frac{S}{\tau_{ph}}+\Gamma g(N-N_{t}) \frac{S}{{(1+\epsilon S)}^b}
\end{eqnarray}

$N$ represents the electron density ($N_{t}$ at transparency) and $S$
the  photon density.  $\tau_{n}$ is the electron spontaneous lifetime
and   $\tau_{ph}$ is  the  photon lifetime. $\beta$  is  the  fraction  of
spontaneous  emission  coupled to the  lasing  mode,  $G$  the
optical confinement factor, $g$ is the differential gain and $\epsilon$
is the gain compression parameter. $q$ is the electron charge,
$V$  the  volume  of the active region and $I$ is the  injection
current which, in general, is a function of time.\\

The  main  novelty in these equations lie in the  non-linear
gain factor that has been traditionally modelled by the term
$g(1-\epsilon S)$, $g/(1+ \epsilon S)$ or even $g/\sqrt{(1+ \epsilon S)}$
\cite{agrawal}. In this  study,
this  term is taken to an arbitrary power \cite{ning} b  in  the
interval  [0,1.5]. In order to cover in the  simulation  the
case  $g(1-\epsilon S)$,  we formally allocate b=-1 to this  case.  We
analyse in this work the effects b has on the static, small-
signal  frequency  response  as  well  as  the  large-signal
temporal variation of the carrier and photon densities.\\

This report is organized as follows: in section 2 we outline
the evaluation of the Laser $S-I$ (photon density versus
injection current) DC characteristics. In section 3, the small-signal frequency response is derived and in section 4, we illustrate the laser response to time-dependent injection currents and section 5 contains our conclusions.

\section{DC CHARACTERISTICS}

The DC limit of the SM laser rate equations is given by:

\begin{eqnarray}
0 & = & \frac{I}{qV}-\frac{N}{\tau_{n}}-g(N-N_{t}) \frac{S}{{(1+\epsilon S)}^b} \\
0 & = & \frac{\Gamma \beta N}{\tau_{n}}-\frac{S}{\tau_{ph}}+\Gamma g(N-N_{t}) \frac{S}{{(1+\epsilon S)}^b}
\end{eqnarray}

From (4) we extract the value of $(N-N_{t})$ as:

\begin{equation}
(N-N_{t})= \left[ \frac { \frac{S}{\tau_{ph}} -\frac{\Gamma \beta N_{t}}{\tau_{n}} }
               {  \Gamma g \frac{S}{{(1+\epsilon S)}^b} + \frac{\Gamma \beta }{\tau_{n}}   } \right]
\end{equation}

and substitute it in (3) relating $I$ to $S$ directly. This
results in:
\begin{equation}
\frac{I}{qV}= \frac{N_{t}}{\tau_{n}} + \frac{(S-\Gamma \beta N_{t})( \frac{1}{\tau_{n}}+ g \frac{S}{{(1+\epsilon S)}^b} )} { \tau_{ph} [ \Gamma g \frac{S}{{(1+\epsilon S)}^b} + \frac{\Gamma \beta }{\tau_{ph}} ]  }
\end{equation}

This relation gives $I$ as a function of $S$. Generating a
series of data values $(S,I)$ by varying $S$ and reverse writing them as
$(I,S)$ will result in the $S-I$ DC characteristic.\\

We consider two laser models A and B (Physical parameters given in Appendix A) and
illustrate this procedure with the characteristics displayed
in Figures 1 and 2 for different values of b. It appears from the
figures that b does not affect significantly the DC characteristics in
contrast with the frequency response and dynamic response as seen in the next sections.

\section{SMALL-SIGNAL FREQUENCY RESPONSE}

In order to derive the small signal frequency response, we
assume all quantities $I$, $N$ and $S$ are taken around some
equilibrium values $I_{0}$, $N_{0}$ and $S_{0}$ and hence:

\begin{equation}
I= I_{0} + \delta I(t), N= N_{0} + \delta N(t), S= S_{0} + \delta S(t)
\end{equation}

This means that equation (1) under variation (7) reads:

\begin{equation}
\frac{d \delta N}{dt}  =  \frac{\delta I}{qV}-\frac{\delta N}{\tau_{n}}-g \delta N \frac{S_{0}}{{(1+\epsilon S_{0})}^b} -g(N_{0}-N_{t}) [ \frac{1}{{(1+\epsilon S_{0})}^b} - \frac{\epsilon S_{0} b}{{(1+\epsilon S_{0})}^{b+1}} ] \delta S \\
\end{equation} 

whereas (2) becomes:

\begin{equation}
\frac{d \delta S}{dt}  =  \frac{\Gamma \beta \delta N}{\tau_{n}}-\frac{\delta S}{\tau_{ph}}+ \Gamma g \delta N \frac{S_{0}}{{(1+\epsilon S_{0})}^b} +\Gamma g(N_{0}-N_{t}) [ \frac{1}{{(1+\epsilon S_{0})}^b} - \frac{\epsilon S_{0} b}{{(1+\epsilon S_{0})}^{b+1}} ] \delta S \\
\end{equation}

In order to tackle the small-signal frequency response we switch to the time-harmonic
case where the time derivatives are given by:
$d\delta I/dt=j\omega \delta I, d\delta N/dt=j\omega \delta N, d\delta S/dt=j\omega \delta S$. This results in a
system of equations relating the three variations $\delta I$, $\delta N$ and
$\delta S$:

\begin{equation}
\delta N  (j\omega +\frac{1}{\tau_{n}} +g \frac{S_{0}}{{(1+\epsilon S_{0})}^b} )= \frac{\delta I}{qV} -g(N_{0}-N_{t}) [ \frac{1}{{(1+\epsilon S_{0})}^b} - \frac{\epsilon S_{0} b}{{(1+\epsilon S_{0})}^{b+1}} ] ) \\
\end{equation}

and:

\begin{equation}
\delta N  ( \frac{\Gamma \beta}{\tau_{n}} +g \frac{ \Gamma S_{0}}{{(1+\epsilon S_{0})}^b} )  =  \delta S ( j\omega + \frac{1}{\tau_{ph}}  -\Gamma g(N_{0}-N_{t}) [ \frac{1}{{(1+\epsilon S_{0})}^b} - \frac{\epsilon S_{0} b}{{(1+\epsilon S_{0})}^{b+1}} ]  \\
\end{equation}

Taking the ratio of the above yields the small-signal frequency response:
\begin{equation}
\frac{\delta S}{\delta I}= \frac{1/E}{C+j\omega G-\omega^2}
\end{equation}

where $E$, $G$ and $C$ are given by:
\begin{equation}
E= \frac{qV}{\left[ \Gamma g \frac{S_{0}}{{(1+\epsilon S_{0})}^b} + \frac{\Gamma \beta }{\tau_{n}}   \right]},
\end{equation}

\begin{equation}
G= \frac{1}{\tau_{n}} + \frac{1}{\tau_{ph}}  -\Gamma g(N_{0}-N_{t}) [ \frac{ 1+\epsilon S_{0}(1-b) }{ { (1+\epsilon S_{0}) }^{b+1} } ] + \frac{g S_{0}}{{(1+\epsilon S_{0})}^{b}},
\end{equation}

\begin{equation}
C= \frac{1}{\tau_{n} \tau_{ph}} - \frac{1}{\tau_{n}}\Gamma g(N_{0}-N_{t})(1-\beta) [ \frac{ 1+\epsilon S_{0}(1-b) }{ { (1+\epsilon S_{0}) }^{b+1} } ] + \frac{1}{\tau_{ph}} \frac{g S_{0}}{{(1+\epsilon S_{0})}^{b}}.
\end{equation}

The standard normalised form ( 0 dB at 0 frequency) of the
frequency response is taken as:

\begin{equation}
\widehat{\frac{\delta S}{\delta I}}= \frac{1}{1+j\omega G/C-\omega^2/C}
\end{equation}

It  is  displayed for laser A and B in Figures 3 and  4  for
different values of b.

In sharp contrast with the DC case, the value of b, has a
drastic effect on the frequency response as observed in the above figures.
We  differ from Way \cite{way} in the frequency response due  to  a
discrepancy  in  the estimation of the resonance  frequency.
Way  defines  the  resonance frequency  as  (using  our  own
notation and adapting it to Way's \cite{way} case):

\begin{equation}
f_{r}=\frac{1}{2\pi} \sqrt{(\frac{\Gamma N_{t} g \tau_{ph}+1}{\tau_{n} \tau_{ph}})} \sqrt{(I/I_{th}-1)}
\end{equation}
            
where $I$ is the bias and $I_{th}$ is the threshold current.
Our corresponding formula by inspection of (16) would be:

\begin{equation}
f_{r}=\frac{1}{2\pi} \sqrt{ \frac{1}{\tau_{n} \tau_{ph}} - \frac{1}{\tau_{n}}\Gamma g(N_{0}-N_{t})(1-\beta) [ \frac{ 1+\epsilon S_{0}(1-b) }{ { (1+\epsilon S_{0}) }^{b+1} } ] + \frac{1}{\tau_{ph}} \frac{g S_{0}}{{(1+\epsilon S_{0})}^{b}} }
\end{equation}            

The  dependence on the bias and threshold currents,  in  our
case,  is  contained in $N_{0}$ and $S_{0}$ that are found numerically
with a Newton method \cite {press} adapted to (5) and (6).\\

Obviously,  Way's estimation \cite{way} of the resonance  frequency  is
approximate  by comparing (17) and (18) since he  wanted  to
get  an  analytical estimate of the resonance frequency.  In
order  to  estimate the discrepancies between our  work  and
Way's  we  display  in Figure 5 the frequency  response  for
various values of the bias current expressed in terms of the
threshold current $I_{th} \sim $21mA.

\section{LARGE-SIGNAL DYNAMIC RESPONSE}

We exploit three possible excitation scheme for the  injection current:
\begin{enumerate}
\item  A step excitation in order to evaluate the step response
of the laser.
\item A graded response with a Gaussian time excitation towards
a higher injection level.
\item  A  modulation  injection  in  order  to  estimate   the
modulation response of the laser for large excursions of the
injection current.
\end{enumerate}

As an example we consider lasers A and B biased at t=0 and
excited with an additional square pulse triggered after 5
nanoseconds of operation. Figures 6-9 show the large signal
dynamic responses for the carrier and photon densities as
functions of time for different values of b. As expected, the
value of b deeply affects the dynamic response of the Laser as 
observed in the figures 6-9. 

\section{CONCLUSIONS}

We  have developed an approach that evaluates the  DC
characteristics, small-signal frequency response and  large-
signal dynamic responses of carrier and photon densities  in
single-mode  semiconductor  laser  diodes directly from the 
rate equations.  The   laser   is described with a pair of
 rate equations containing  a  novel non-linear  gain  compensation term depending  on  a  single parameter  b  that can be chosen arbitrarily  in  the  range [0,1.5]  as  has  been  shown recently  with  a  microscopic
calculation  of  plasma-heating induced  intensity-dependent
gain effects \cite{ning}.\\

Our DC evaluations agree with several published results \cite{agrawal,ning,darcie,way,bowers86,bowers87} 
and our large-signal dynamical reponses  agree also with
what is well established in the literature \cite{channin}. We differ
with the small-signal frequency response results given by
Way \cite{way} due to a discrepancy in the frequency response and
resonance frequency estimation.\\

The parameter b has almost no effect on the DC characteristics but deeply
affects the small-signal frequency response as well as the large-signal dynamic
response.

The methodology developed in this work can be easily
generalised to an arbitrary number of state equations that
appear in multimode semiconductor lasers, MQW (Muli-Quantum-
Well) or Strained layer lasers \cite{rideout,tessler,makino}.\\

{\bf Acknowledgement} \\
This work started while the author was with the Department of  Electrical Engineering 
and with TRLabs in Saskatoon, Canada. The  author wishes to acknowledge friendly discussions with David Dodds regarding some aspects of the problem. This work was supported in part by a Canada NSERC University fellowship
grant.

\newpage

\vspace{1cm}
\centerline{\Large\bf Appendix}
\vspace{1cm}

\begin{table}[htbp]
\begin{center}
\begin{tabular}{ ||c|c||}
\hline
 \multicolumn{2}{||c||}{LASER A (ORTEL LS-620)}\\
\hline
$G$ (mode confinement factor)    &          0.646  \\
\hline
  $\tau_{n}$ (electron spontaneous   lifetime)    &         $3.72 \hspace{1mm} 10^{-9} \hspace{1mm} sec$ \\
\hline       
 $\tau_{ph}$ (photon lifetime)     &            $ 2. 10^{-12} sec$ \\
\hline
 $N_{t}$ (electron density at  transparency)    &       $4.6 \hspace{1mm}10^{24} m^{-3}$  \\
\hline     
$g$ (differential gain )        &     $10^{-12} \hspace{1mm} m^{3}/sec$  \\
\hline
$\epsilon$ (gain compression   parameter)     &         $ 3.8 \hspace{1mm} 10^{-23}\hspace{1mm} m^{3}$ \\
\hline
 $\beta $ (fraction of spontaneous emission coupled to the lasing mode)   &         0.001 \\
\hline
  $V$ (volume of the active region)      &      $9. 10^{-17} m^{3}$ \\
 \hline
\end{tabular}
\caption{Physical parameters of Laser A}
\end{center}
\label{tab1}
\end{table}

\begin{table}[htbp]
\begin{center}
\begin{tabular}{ ||c|c||}
\hline
 \multicolumn{2}{||c||}{LASER B}\\
\hline
$G$ (mode confinement factor)    &          0.34  \\
\hline
  $\tau_{n}$ (electron spontaneous   lifetime)    &         $3. \hspace{1mm} 10^{-9} \hspace{1mm} sec$ \\
\hline       
 $\tau_{ph}$ (photon lifetime)     &            $ 2. 10^{-12} sec$ \\
\hline
 $N_{t}$ (electron density at  transparency)    &       $10^{24} m^{-3}$  \\
\hline     
$g$ (differential gain )        &     $3. 10^{-12} \hspace{1mm} m^{3}/sec$  \\
\hline
$\epsilon$ (gain compression   parameter)     &         $ 3. \hspace{1mm} 10^{-23}\hspace{1mm} m^{3}$ \\
\hline
 $\beta $ (fraction of spontaneous emission coupled to the lasing mode)   &         0.001 \\
\hline
  $V$ (volume of the active region)      &      $3.6 \hspace{1mm} 10^{-18} m^{3}$ \\
 \hline
\end{tabular}
\caption{Physical parameters of Laser B}
\end{center}
\label{tab2}
\end{table}


\newpage

\vspace{1cm}
\centerline{\Large\bf Figure Captions}
\vspace{1cm}

\begin{itemize}

\item[Fig.\ 1:] $S-I$ characteristics (Laser A) for various  values
of  b: -1.0, 0.5, 1.0 and 1.5. The different values of b  do
not affect significantly the DC characteristics.

\item[Fig.\ 2:] $S-I$ characteristics (Laser B) for various  values
of  b: -1.0, 0.5, 1.0 and 1.5. The different values of b  do
not affect significantly the DC characteristics.

\item[Fig.\ 3:] Small signal frequency response amplitude  (Laser
A)  versus frequency for various values of b: -1.0, 0.5, 1.0
and 1.5. The bias current is chosen to be 40 mA.

\item[Fig.\ 4:] Small signal frequency response amplitude  (Laser
B)  versus frequency for various values of b: -1.0, 0.5, 1.0
and 1.5. The bias current is chosen to be 1 mA.

\item[Fig.\ 5:] Small signal frequency response amplitude (Laser A
or  Way's  case [4]) versus frequency for various values  of
the  bias  current. The bias current is taken as  $I_{th}$,  1.25
$I_{th}$, 1.75 $I_{th}$ and 2.5 $I_{th}$ where $I_{th} \sim$ 21mA.

\item[Fig.\ 6:] Laser A large signal dynamic response amplitude of
the  carrier density versus time for various values of b:
-1.0,  0.5, 1.0. The bias current is 40 mA and a square pulse
excitation  of  10  mA is applied after 5  nanoseconds.  The
dynamic  response for b=1.5 is off the graph scale.

\item[Fig.\ 7:] Laser A large signal dynamic response amplitude of
the  photon density versus time for various values of  b:
-1.0,  0.5, 1.0. The bias current is 40 mA and a square pulse
excitation  of  10  mA is applied after 5  nanoseconds.  The
dynamic  response for b=1.5 is off the graph scale.

\item[Fig.\ 8:] Laser B large signal dynamic response amplitude of
the  carrier density versus time for various values of b:
-1.0,  0.5, 1.0. The bias current is 1 mA and a square  pulse
excitation  of  0.5 mA is applied after 5  nanoseconds.  The
dynamic  response for b=1.5 is  off the graph scale.

\item[Fig.\ 9:] Laser B large signal dynamic response amplitude of
the  photon density versus time for various values of  b:
-1.0,  0.5, 1.0. The bias current is 1 mA and a square  pulse
excitation  of  0.5 mA is applied after 5  nanoseconds.  The
dynamic  response for b=1.5 is  off the graph scale.

\end{itemize}

\newpage

\begin{figure}[hbp]
\begin{center}
\scalebox{0.8}{\includegraphics{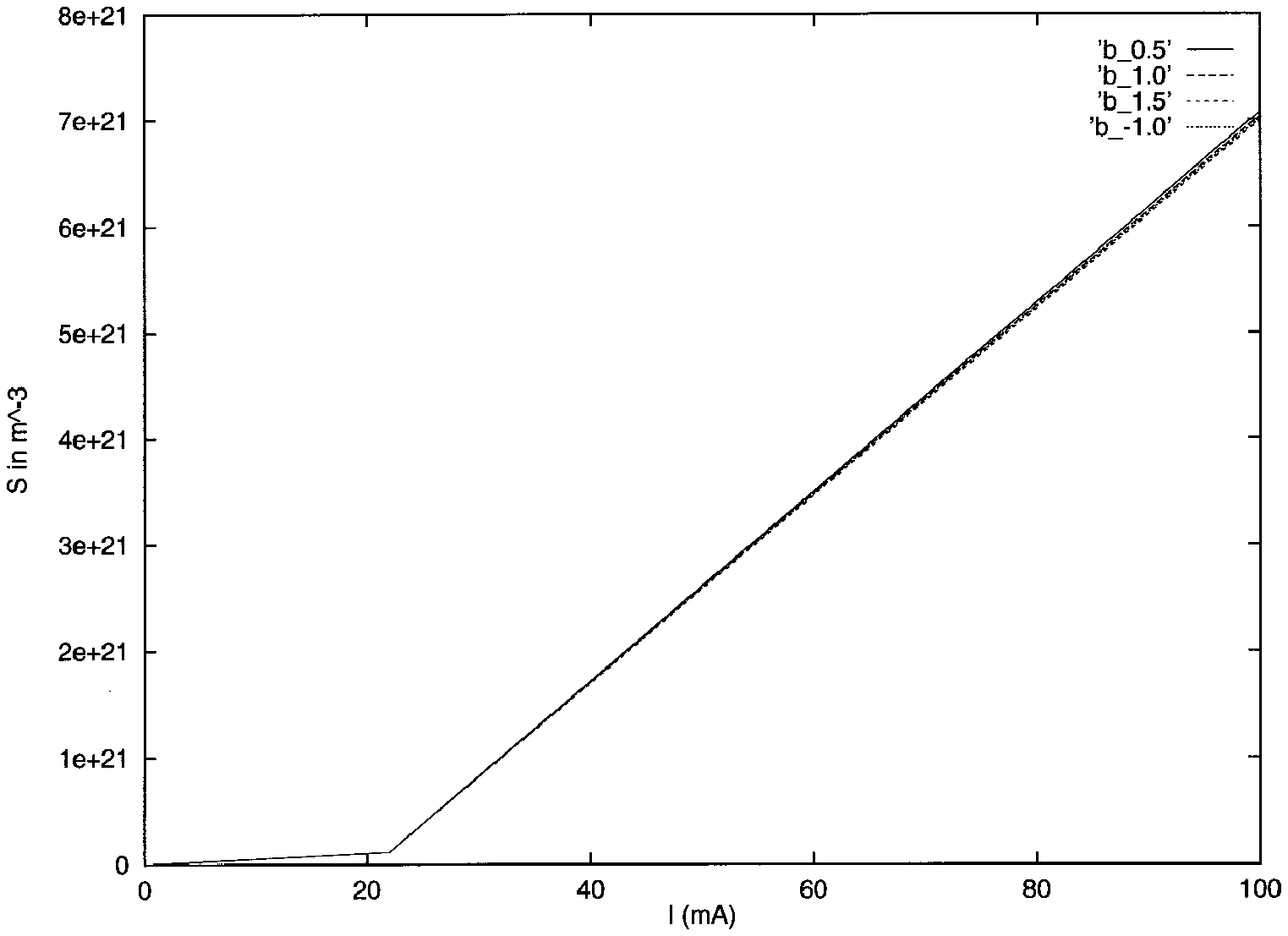}}
\end{center}
\caption{$S-I$ characteristics (Laser A) for various  values
of  b: -1.0, 0.5, 1.0 and 1.5. The different values of b  do
not affect significantly the DC characteristics.} \label{fig1}
\end{figure}
\eject

\begin{figure}[hbp]
\begin{center}
\scalebox{0.8}{\includegraphics{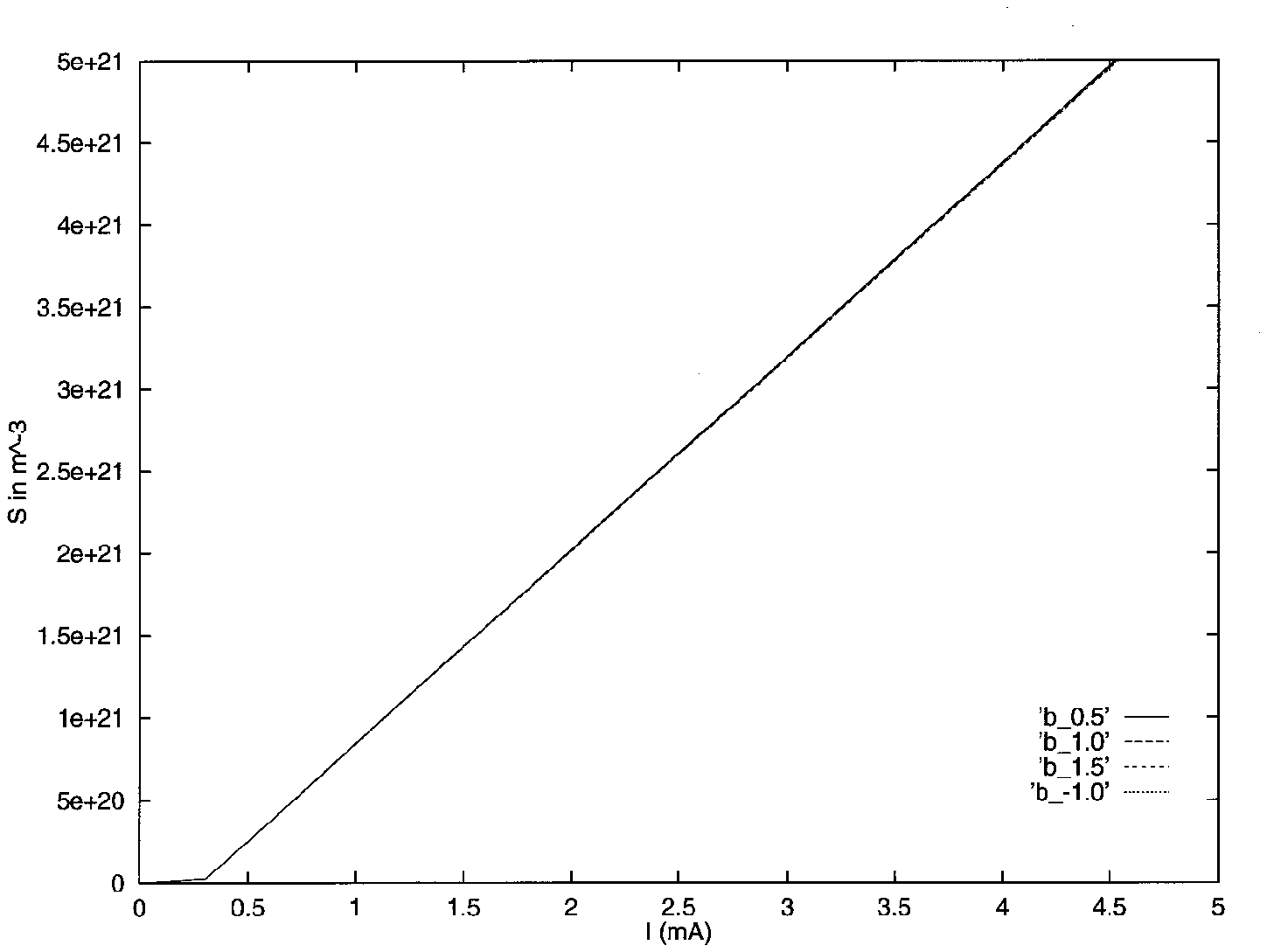}}
\end{center}
\caption{$S-I$ characteristics (Laser B) for various  values
of  b: -1.0, 0.5, 1.0 and 1.5. The different values of b  do
not affect significantly the DC characteristics.} \label{fig2}
\end{figure}
\eject

\begin{figure}[hbp]
\begin{center}
\scalebox{0.8}{\includegraphics{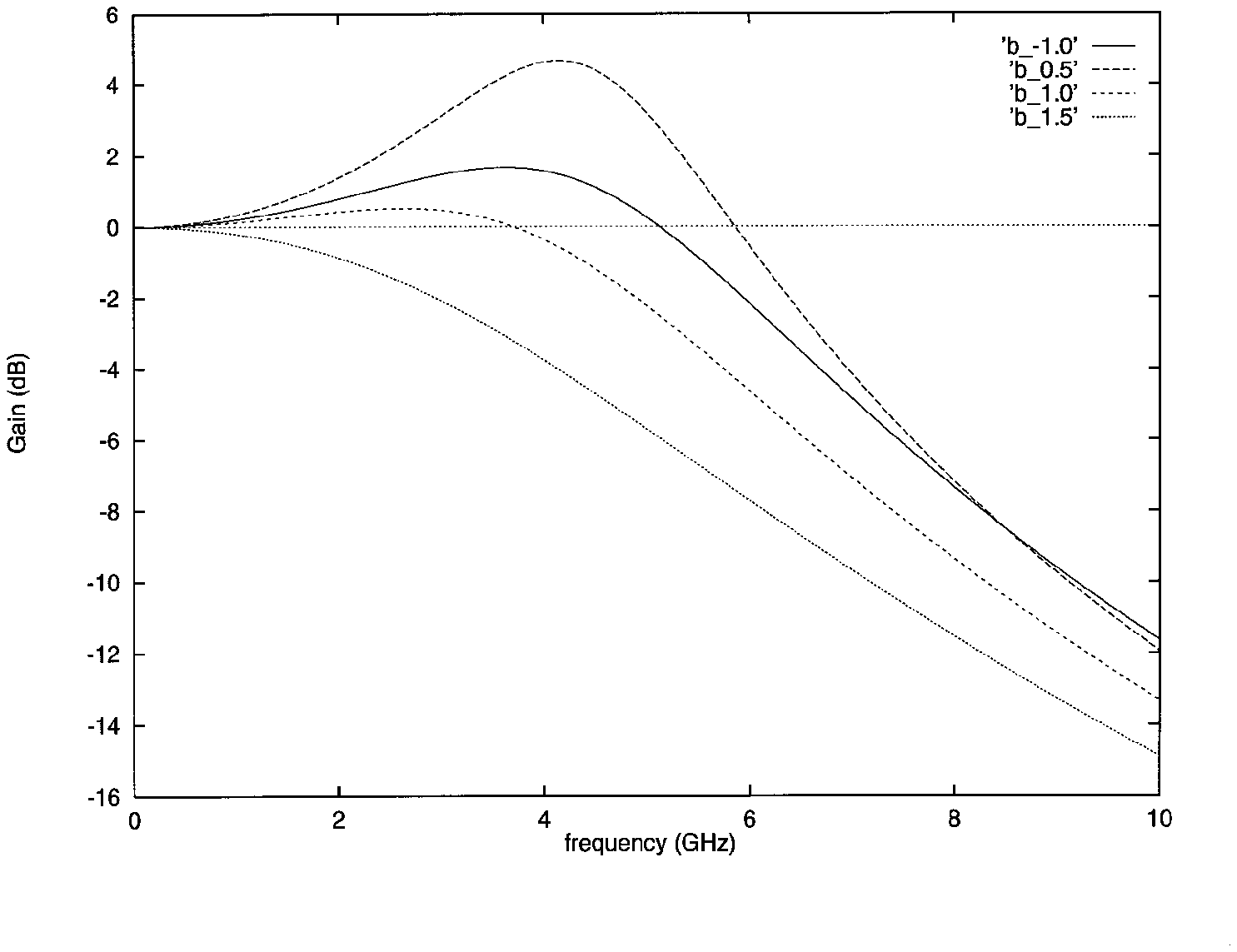}}
\end{center}
\caption{Small signal frequency response amplitude  (Laser
A)  versus frequency for various values of b. From top to bottom, b is -1.0, 0.5, 1.0
and 1.5 respectively. The bias current is chosen as 40 mA.} \label{fig3}
\end{figure}
\eject

\begin{figure}[hbp]
\begin{center}
\scalebox{0.8}{\includegraphics{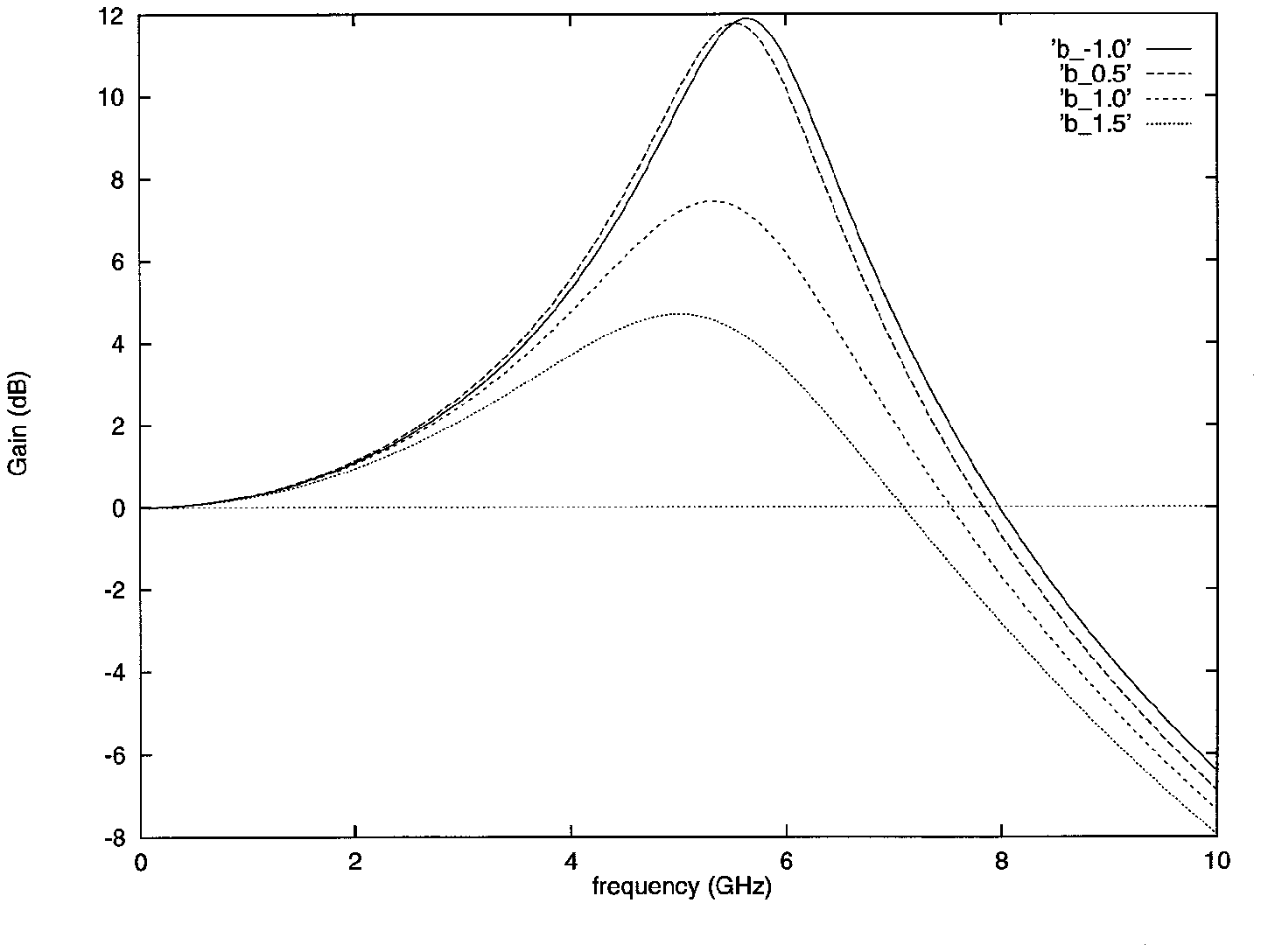}}
\end{center}
\caption{Small signal frequency response amplitude  (Laser
B)  versus frequency for various values of b. From top to bottom, b is -1.0, 0.5, 1.0
and 1.5 respectively. The bias current is chosen as 1 mA.} \label{fig4}
\end{figure}
\eject

\begin{figure}[hbp]
\begin{center}
\scalebox{0.8}{\includegraphics{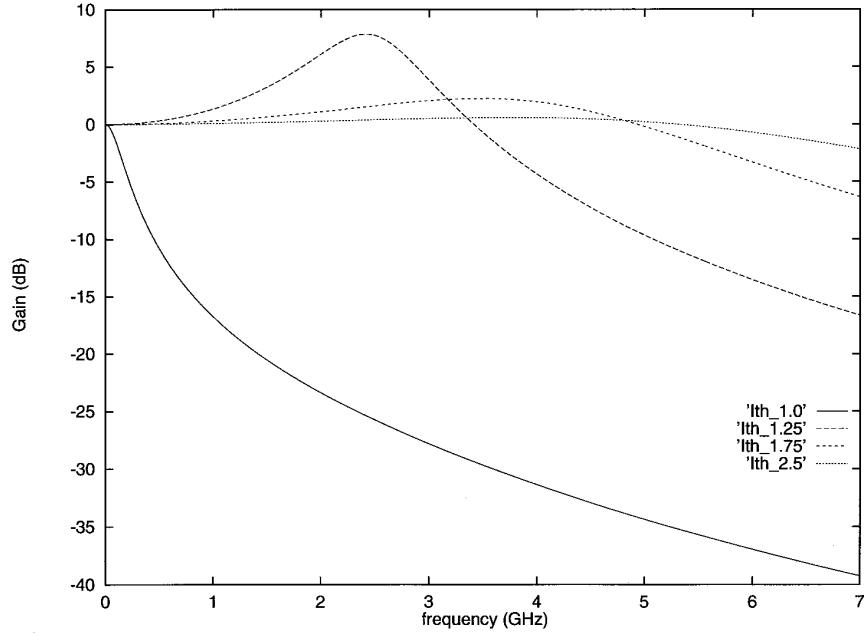}}
\end{center}
\caption{Small signal frequency response amplitude (Laser A
or  Way's  case [4]) versus frequency for various values  of
the  bias  current. From top to bottom, the bias current is taken as  $I_{th}$,  1.25
$I_{th}$, 1.75 $I_{th}$ and 2.5 $I_{th}$ where $I_{th} \sim$ 21mA.} \label{fig5}
\end{figure}
\eject

\begin{figure}[hbp]
\begin{center}
\scalebox{0.8}{\includegraphics{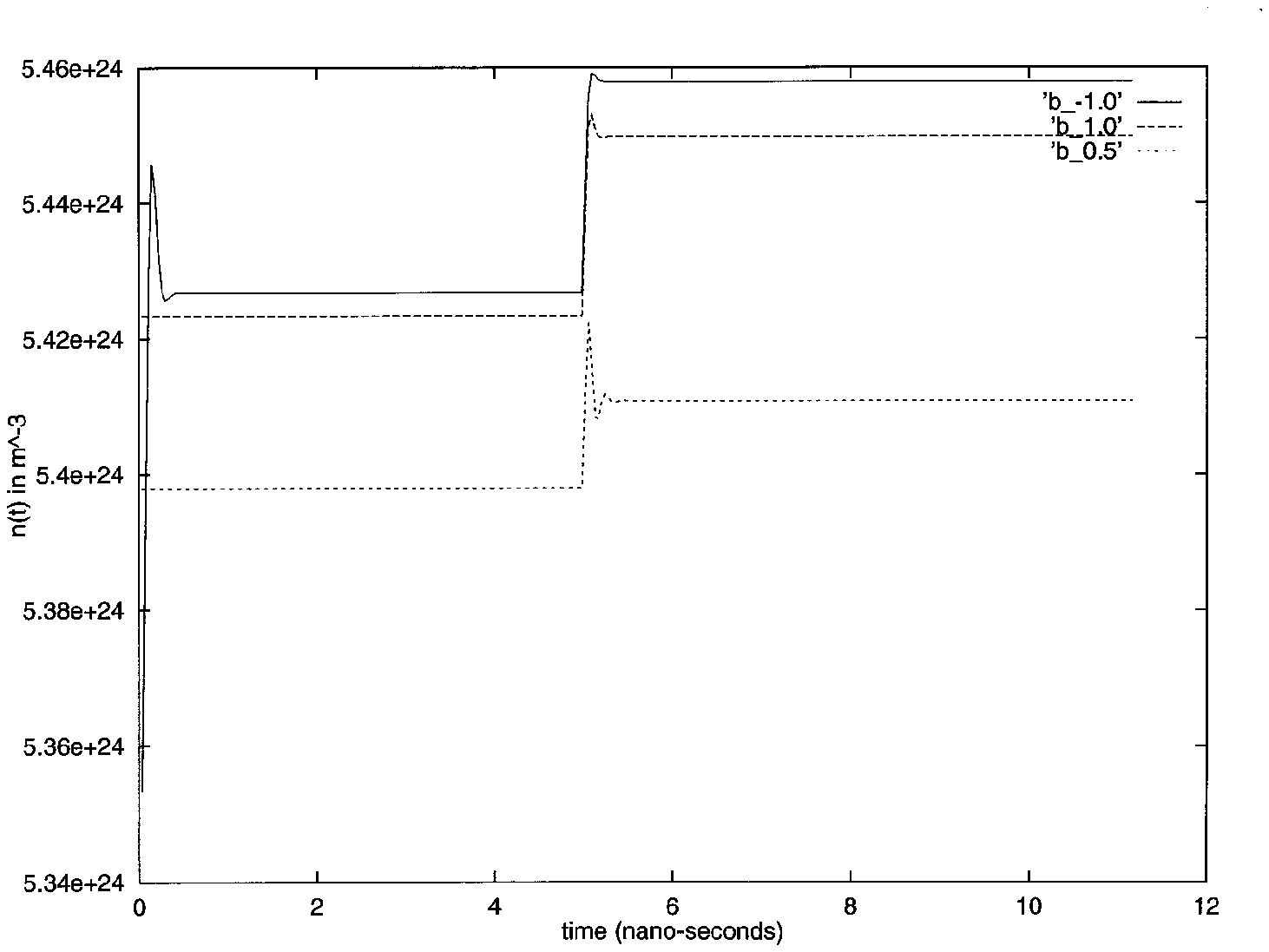}}
\end{center}
\caption{Laser A large signal dynamic response amplitude of
the  carrier density versus time for various values of b:
-1.0,  0.5, 1.0. The bias current is 40 mA and a square pulse
excitation  of  10  mA is applied after 5  nanoseconds.  The
dynamic  response for b=1.5 is  off the graph scale.} \label{fig6}
\end{figure}
\eject

\begin{figure}[hbp]
\begin{center}
\scalebox{0.8}{\includegraphics{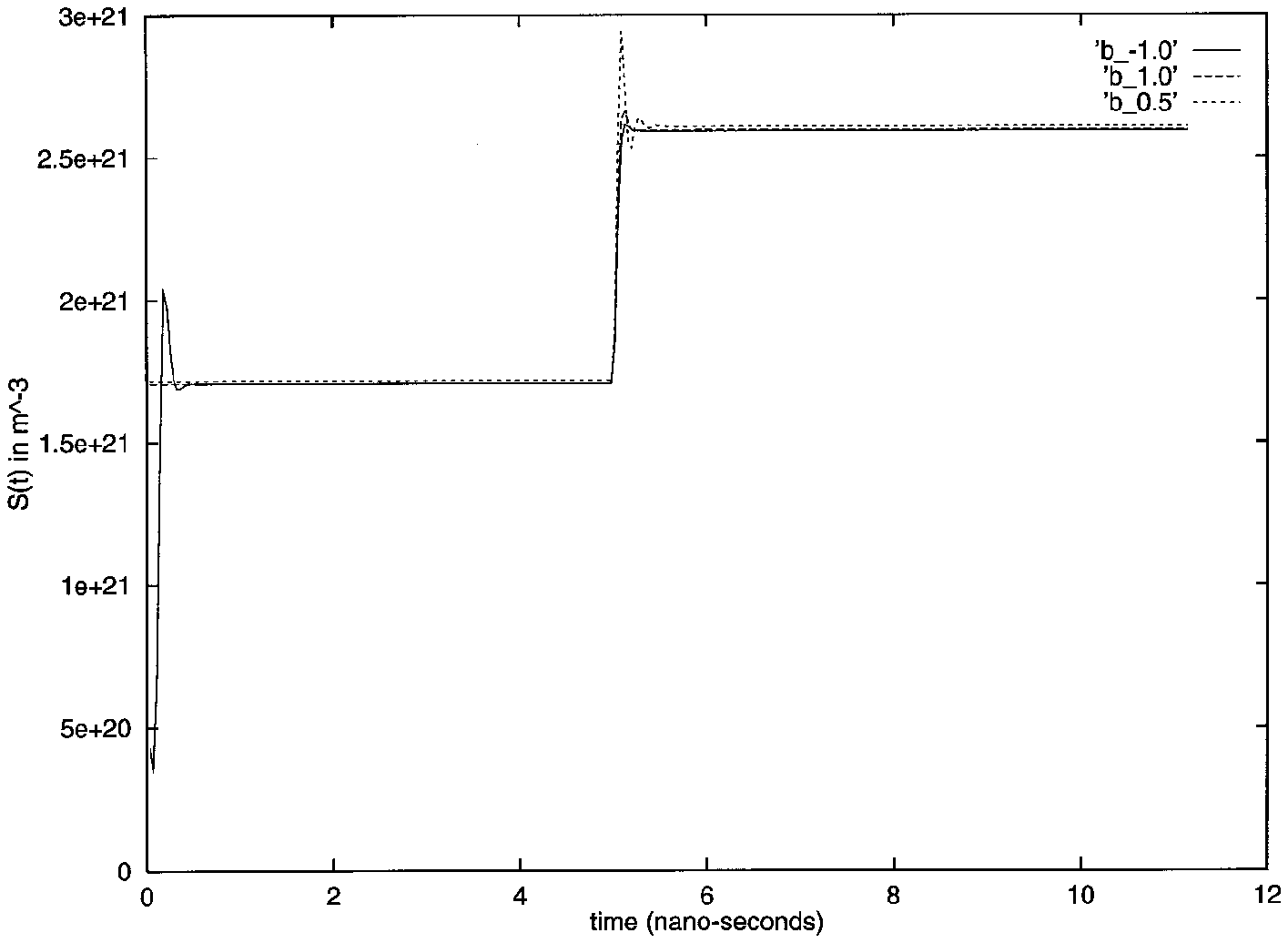}}
\end{center}
\caption{Laser A large signal dynamic response amplitude of
the  photon density versus time for various values of  b:
-1.0,  0.5, 1.0. The bias current is 40 mA and a square pulse
excitation  of  10  mA is applied after 5  nanoseconds.  The
dynamic  response for b=1.5 is  off the graph scale.} \label{fig7}
\end{figure}
\eject

\begin{figure}[hbp]
\begin{center}
\scalebox{0.8}{\includegraphics{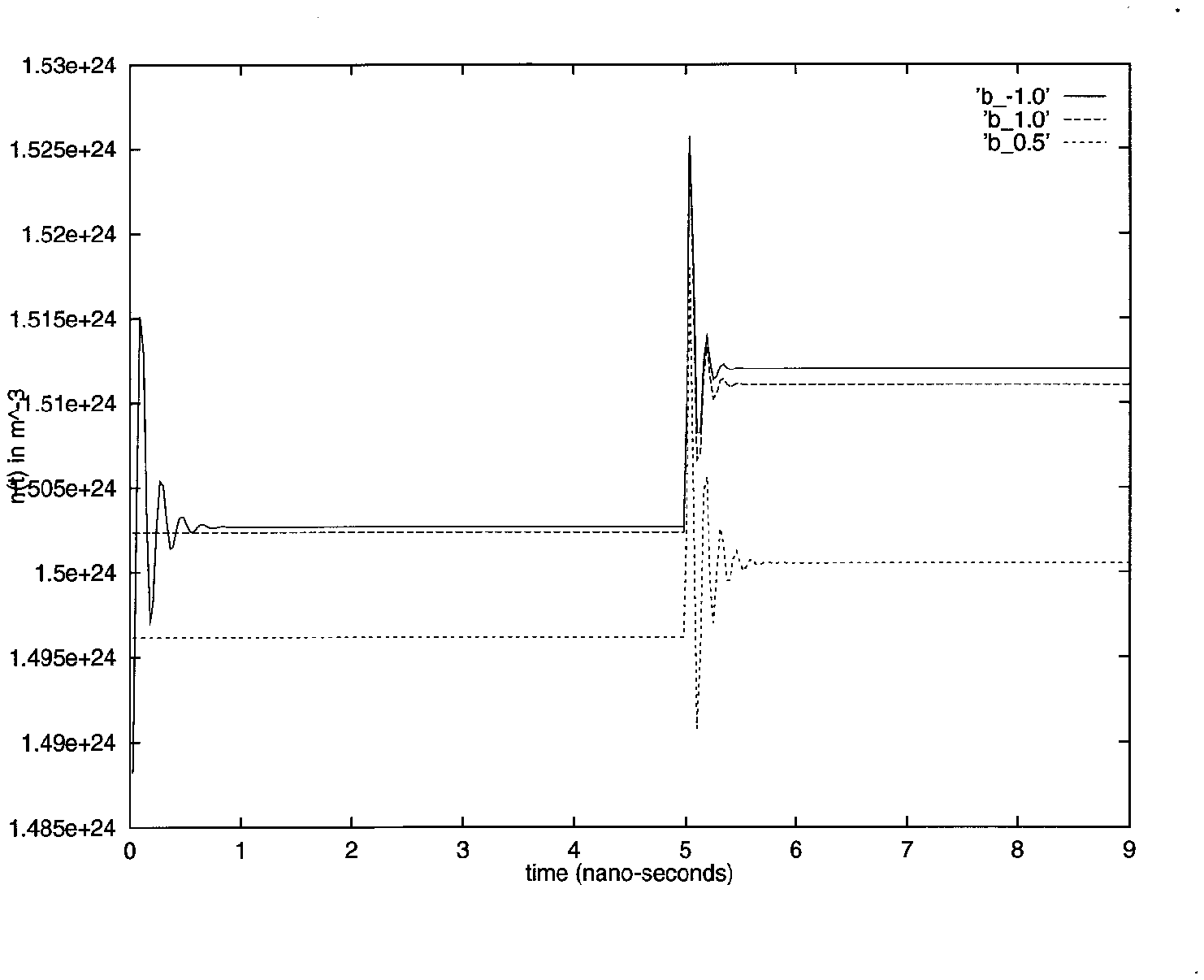}}
\end{center}
\caption{Laser B large signal dynamic response amplitude of
the  carrier density versus time for various values of b:
-1.0,  0.5, 1.0. The bias current is 1 mA and a square  pulse
excitation  of  0.5 mA is applied after 5  nanoseconds.  The
dynamic  response for b=1.5 is off the graph scale.} \label{fig8}
\end{figure}
\eject

\begin{figure}[hbp]
\begin{center}
\scalebox{0.8}{\includegraphics{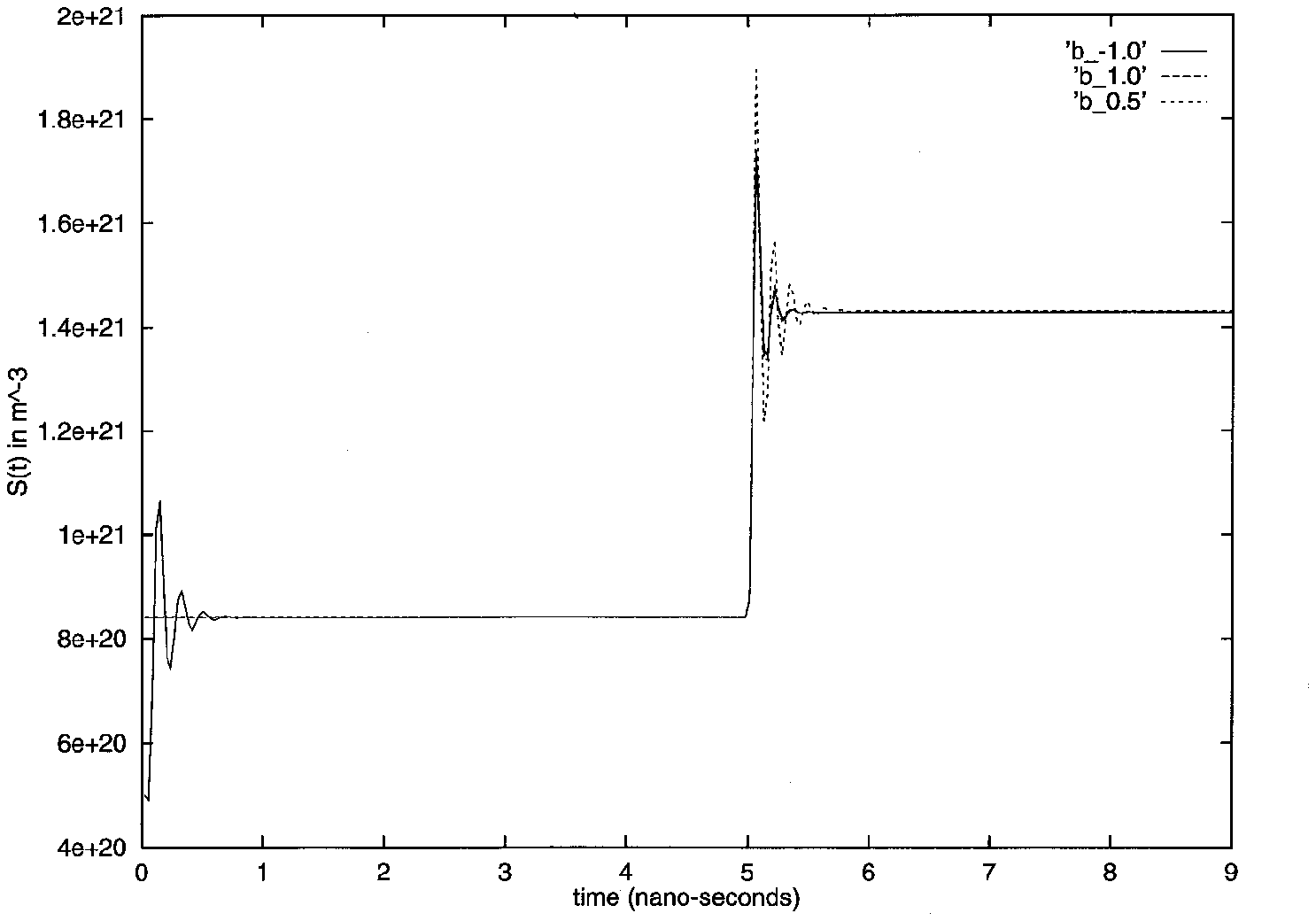}}
\end{center}
\caption{Laser B large signal dynamic response amplitude of
the  photon density versus time for various values of  b:
-1.0,  0.5, 1.0. The bias current is 1 mA and a square  pulse
excitation  of  0.5 mA is applied after 5  nanoseconds.  The
dynamic  response for b=1.5 is off the graph scale.} \label{fig9}
\end{figure}

\end{document}